\documentclass[
  aip,
  apl,
  amsmath,amssymb,
  floatfix,
  reprint,
]{revtex4-2}

\usepackage{graphicx}
\usepackage{dcolumn}
\usepackage{bm}
\usepackage[utf8]{inputenc}
\usepackage[T1]{fontenc}
\usepackage{mathptmx}
\usepackage{etoolbox}
\usepackage{hyperref}

\makeatletter
\def\@email#1#2{%
 \endgroup
 \patchcmd{\titleblock@produce}
  {\frontmatter@RRAPformat}
  {\frontmatter@RRAPformat{\produce@RRAP{*#1\href{mailto:#2}{#2}}}\frontmatter@RRAPformat}
  {}{}
}%
\makeatother

\begin{document}


\title{Compressibility of micromagnetic solutions in tensor train format}

\author{Thierry Valet}
\email{tvalet@mphysx.com}
\affiliation{MPhysX O\"U, Tallinn, Estonia}
\affiliation{AIMR and RIEC, Tohoku University, Sendai, Japan}

\author{Nicolas Vukadinovic}
\affiliation{CEA, DAM, Le Ripault, F-37260 Monts, France}

\date{\today}

\begin{abstract}
For three‑dimensional (3D) magnetic objects with linear size $L$ exceeding a few exchange lengths, the micromagnetic state exhibits pronounced informational sparsity: low‑dimensional, high‑gradient regions (e.g., domain walls) coexist with near‑uniformly magnetized volumetric domains. Because standard micromagnetic simulation methods discretize the magnetization on near‑uniform 3D grids with linear cell size $a$, they cannot take advantage of this sparsity. The computational problem scales as $\sim L^3$ and $\sim (1/a)^3$. In this Letter, we establish that direct tensor-train (TT) representations overcome these poor scalings by exploiting the spatial sparsity optimally, while preserving accuracy in a controlled way. Focusing on representative flux-closure configurations in soft-magnetic rectangular prisms, in the near-micrometer regime, we demonstrate that the parameter count of TT-compressed micromagnetic data scales approximately as $L^{1.8\text{--}1.9}$ and $(1/a)^{1.2\text{--}1.3}$. Hence the relative advantage over dense discretizations rapidly {\em grows} with the problem size and refinement level. These first results provide a strong motivation for future developments of micromagnetic solvers in TT format which could transcend the limitations of traditional simulators, with far reaching potential impacts on fundamental research and technology development.
\end{abstract}

\maketitle

\section{Introduction}
\label{sec:introduction}

Numerical simulations based on the micromagnetic theory \cite{brown:1963} are essential for a deep understanding of the static and dynamic properties of magnetic nanostructures and devices in the fundamentally and technologically relevant fields of magnonics and spintronics  \cite{abert:2019,trabzon:2025}. Standard numerical methods require a dense spatial discretization at the level of the usually nanometric physical length scale {e.g.}, the exchange length $l_{\rm ex}$, which controls the emergence of key spatial structures such as domain walls. Current trends in research and technology often require to simulate fully three-dimensional (3D) magnetic objects with potentially one or more linear dimensions $L$ exceeding one micrometer \cite{Fernandez:2017,Gu:2025}. With standard micromagnetics simulators \cite{mumax3,oommf}, the inherent data structure scaling as \smash{$\sim (L/l_{\rm ex})^3$,} compounded by the need to simulate many configurations across materials parameters, geometries and possibly time \cite{abert:2019}, leads to insurmountable computational bottlenecks.

To address similar challenges associated with multidimensional data \cite{de:2024,li:2025} and discretized multivariate functions \cite{ritter:2024}, low‑rank factorization in the tensor‑train (TT) format \cite{oseledets:2011} has emerged as a powerful approach. The tensor cross‑interpolation (TCI) algorithm \cite{oseledets:2010,dolgov:2020} is particularly attractive in this context,  as it constructs the TT factorizations with controlled accuracy and only requires sparse sampling of the initial data. In the somewhat related context of computational fluid dynamics, recent studies have shown that TT representations can achieve orders‑of‑magnitude compression of simulation data with prescribed error tolerance \cite{pisoni:2025,danis:2025}. Micromagnetic configurations are naturally information‑sparse in space, with large near-uniformly magnetized volumetric domains interspersed with high‑gradient structures of reduced dimensionality {\em e.g.}, domain walls. This spatial sparsity strongly suggests that compact TT representations may exist, however and to the best of our knowledge, a systematic study of micromagnetic compressibility in TT format has yet to be reported.

In this Letter, we address this gap with a systematic study of TT-TCI compression of simulated micromagnetic equilibrium configurations. We focus on a deliberately simple yet physically nontrivial and representative system: an isolated, soft‑magnetic, rectangular  prism in a flux‑closure state, with linear dimensions ranging from one hundred nanometers to nearly one micrometer. The underlying micromagnetic physics is not in question. Equilibrium flux-closure states are known to be organized by localized walls, edge-turning regions, and other structures of reduced effective dimensionality. The question is whether a direct TT compression can exploit that organization efficiently enough for the compressed complexity to scale with those high-gradient region rather than with the full volume. We first examine how the TT-compressed size grows with the prism size at fixed sampling pitch, and then how it grows under grid refinement at fixed physical size. The resulting scalings, respectively near-quadratic and near-linear, demonstrate that direct TT compression does in fact capture the spatially non-extensive organization of 3D micromagnetic configurations in an optimal way. We conclude by discussing how these very favorable scaling opens a new path to transcend the limits of established numerical methods.

\section{Methods}
\label{sec:methods}
All micromagnetic simulations, TT compressions and statistical data analysis reported in this work were performed on a compute node running Ubuntu 24.04.4 LTS (Noble Numbat). The computational resources comprised a single Intel Xeon Platinum 8259CL processor clocked at 2.50 GHz, with 8 physical cores and 16 logical CPUs, together with 60 GB of system memory. Hardware acceleration was provided by one NVIDIA Tesla T4 GPU equipped with 15 GB of on-board device memory.

\subsection{Micromagnetic simulations}
\label{sec:methods_micromagnetic}
The equilibrium micromagnetic solutions analyzed in this work were computed with a python scripted and GPU accelerated simulation pipeline based on the \texttt{mumaxplus}\cite{Moreels2026} package.
The considered prototype system is an isolated rectangular prism of aspect ratio $2{:}1{:}1$ in which every simulation cell is magnetic. Hence, the magnetic body coincide with the full simulation domain.
Periodic boundary conditions were disabled in all three directions, resulting in a demagnetizing field consistent with a finite magnetized volume in free space. The used set of material parameters are those of a typical soft-magnetic material, with saturation magnetization $M_{\mathrm{s}}=8.0\times10^5\,\mathrm{A/m}$, exchange stiffness $A=1.3\times10^{-11}\,\mathrm{J/m}$ and vanishing anisotropy. This corresponds to an exchange length $l_{\mathrm{ex}}=\sqrt{2A/(\mu_0 M_{\mathrm{s}}^2)}\approx5.69\,\mathrm{nm}$, defining the characteristic length scale of the localized structures. The attained micromagnetic equilibrium depends on the initial configuration in the explored size range, since typically multiple local minima of the micromagnetic energy coexist. Consequently, all cases were initialized from the same analytic three-dimensional flux-closure ansatz.
The chosen initial configuration combines a dominant in-plane circulation around the prism long  {\em i.e.}, the $x$ axis, with weaker auxiliary curls in the $yz$ and $zx$ planes, an inward component near the top and bottom surfaces, and a weak deterministic symmetry-breaking perturbation.
Each case was then evolved to a local energy minimum with the \texttt{mumaxplus} built-in \texttt{Minimize} function, using a stopping tolerance of $10^{-6}$.
For all considered cases, this protocol converged to the same family of nontrivial 3D solutions \cite{Yan:2007}, as illustrated in Fig.~\ref{fig:benchmark_render}: extended closure domains, symmetric surface vortices at the top and bottom $xy$ planes, with clearly localized domain walls and edge-turning regions. These different localized structures carry most of the strong spatial variations. Two series of micromagnetic simulations were performed.
The first one was a variable-size (VS) series, with the cell size kept fixed at $a=1.5625\,\mathrm{nm}$, corresponding to $l_{\mathrm{ex}} / a \approx 3.638$, and with increasing physical dimensions at constant aspect ratio, ranging from $100\times50\times50\,\mathrm{nm}^3$ to $600\times300\times300\,\mathrm{nm}^3$. The second one was a variable refinement (VR) series, in which the physical prism dimensions were kept constant at $200\times100\times100\,\mathrm{nm}^3$, while the grid varied as $(N_x,N_y,N_z)=(N_x,N_x/2,N_x/2)$ with $N_x=96, 128, 160, 192, 224$, and $256$. The resulting physically relevant rounded sampling ratios $l_{\mathrm{ex}}/a$ are $2.729$, $3.638$, $4.548$, $5.457$, $6.367$ and $7.276$, respectively.

\begin{figure}[t]
  \centering
  \includegraphics[width=0.49\textwidth]{figs/fig_0.png}
  \caption{\textbf{Representative equilibrium flux-closure state for the considered magnetic rectangular prisms.} We use a surface line-integral convolution rendering to reveal the in-plane magnetization flow at the surface of the prism, and a color scale to indicate the amplitude of the z component. Closure domains and a surface vortex are clearly visible, highlighting the non‑trivial and fully three‑dimensional micromagnetic configuration.}
  \label{fig:benchmark_render}
\end{figure}

\subsection{Tensor-train compression}
\label{sec:methods_tt}
Each converged micromagnetic solution \(\mathbf{m}\) was stored as a dense fourth-order tensor $\mathcal{M}_{c, i, j, k}\in\mathbb{R}^{3\times N_x\times N_y\times N_z}$, where the first index labels the
Cartesian magnetization component, and remaining one index the grid points in each spatial directions. These fourth-order tensors were compressed in a TT format \cite{oseledets:2010}, as a product of third-order TT cores, according to
\begin{equation}
    \mathcal{M}_{c, i, j, k} \approx \mathcal{M}_{c, i, j, k}^{\mathrm TT} = \mathcal{C}_{\scriptscriptstyle (1)}{}_{1, c}^{\alpha_1} \mathcal{C}_{\scriptscriptstyle (2)}{}_{\alpha_1, i}^{\alpha_2} \mathcal{C}_{\scriptscriptstyle (3)}{}_{\alpha_2, j}^{\alpha_3} \mathcal{C}_{\scriptscriptstyle (4)}{}_{\alpha_3, k, 1},
\end{equation}
with implicit summation over the repeated internal dummy (bond) indices $\alpha_\mu$ connecting adjacent cores. The maximum size of the bond indices for a given TT is its {\em rank}. These factorization were numerically performed using the TCI algorithm as implemented in the \texttt{tntorch}\cite{UBS:22} Python package, running on \texttt{PyTorch}\cite{paszke2019pytorch} (version 2.6.0), with GPU acceleration. The compressed parameter count (CPC) is defined as the total number of stored entries in the TT, and the compression ratio (CR) as the ratio between the CPC and the initial dense tensor entry count.
Reconstruction error is defined as the maximal pointwise vector error (MPVE) over the full grid {\em i.e.}, \smash{$\max_{i,j,k}\|{\bf m}_{\mathrm{TT}}-{\bf m}\|_2$}, with ${\bf m}_{\mathrm{TT}}$ the reconstructed magnetization vector from the TT factorization. We chose this metric since it is local and sensitive to reconstruction errors likely to concentrate in high gradient regions. In order to derive curves establishing the CR variation as a function of the achieved MPVE, for a given initial dense tensor, multiple TCI were performed with increasing rank constraint, together with a stringent in-sample tolerance and comfortable maximum iteration budget, insuring that the rank was always the actual stopping criteria for the TCI. Owing to the sensitivity of the TCI algorithm to the initial choice of pivot, five different random seeds were used for each compression series as a function of rank, for a given dense tensor. This allowed for estimating mean variation and confidence intervals (CI). For estimating the CPC power scaling laws, all rank-sweep records in the common window $10^{-3}\leq \mathrm{MPVE}\leq 5\times10^{-2}$ were fitted separately for the VS and VR series, using Gaussian generalized estimating equations with exchangeable within-case correlation. This was implemented with \texttt{pandas}\cite{mckinney-proc-scipy-2010} and \texttt{statsmodels}\cite{seabold2010statsmodels}; target-adjusted CPC scaling lines and $95\%$ CI on the fitted exponents were then extracted from these models.

\begin{figure*}[!t]
  \centering
  \includegraphics[width=0.49\textwidth]{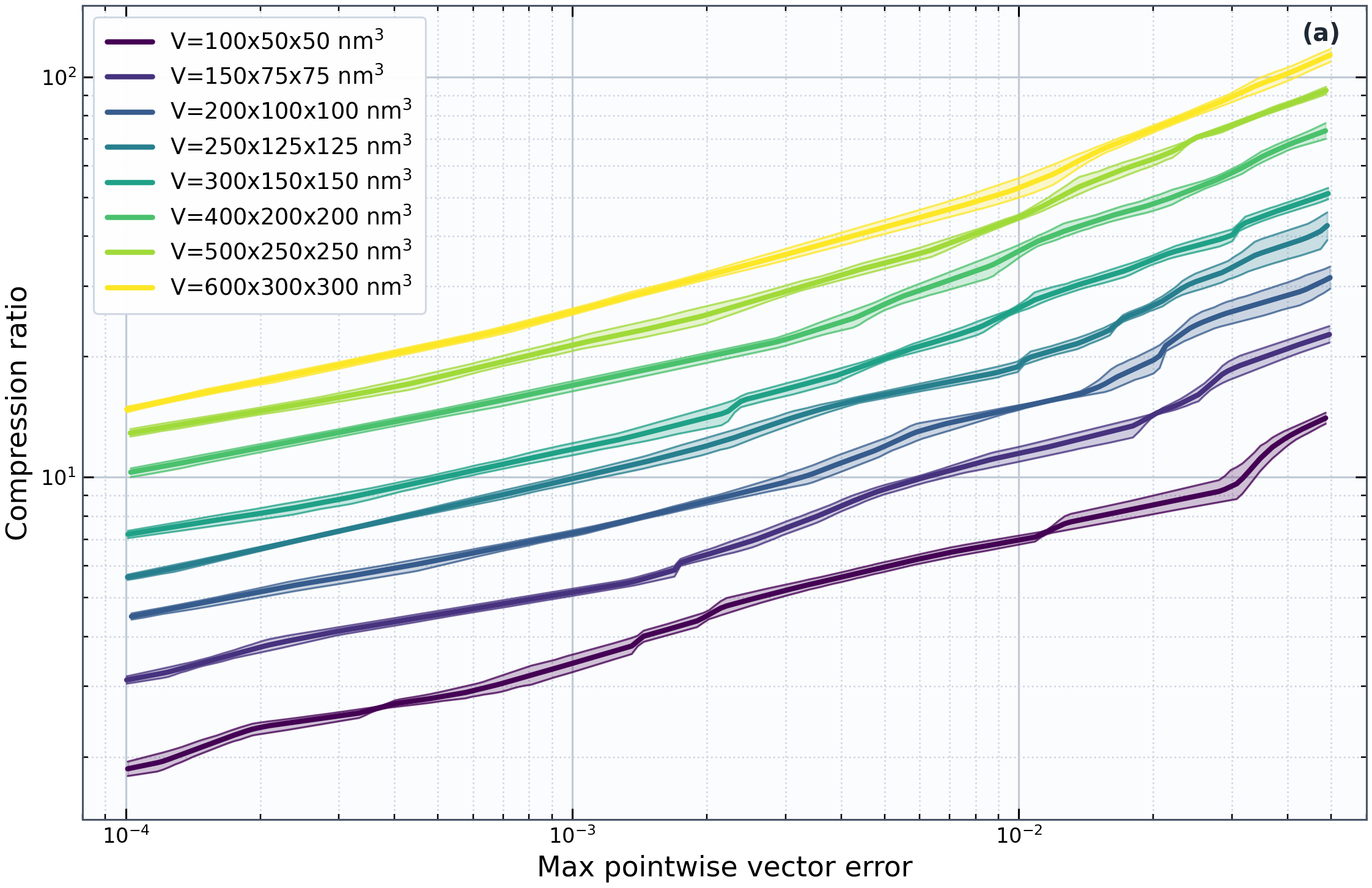}\hfill
  \includegraphics[width=0.49\textwidth]{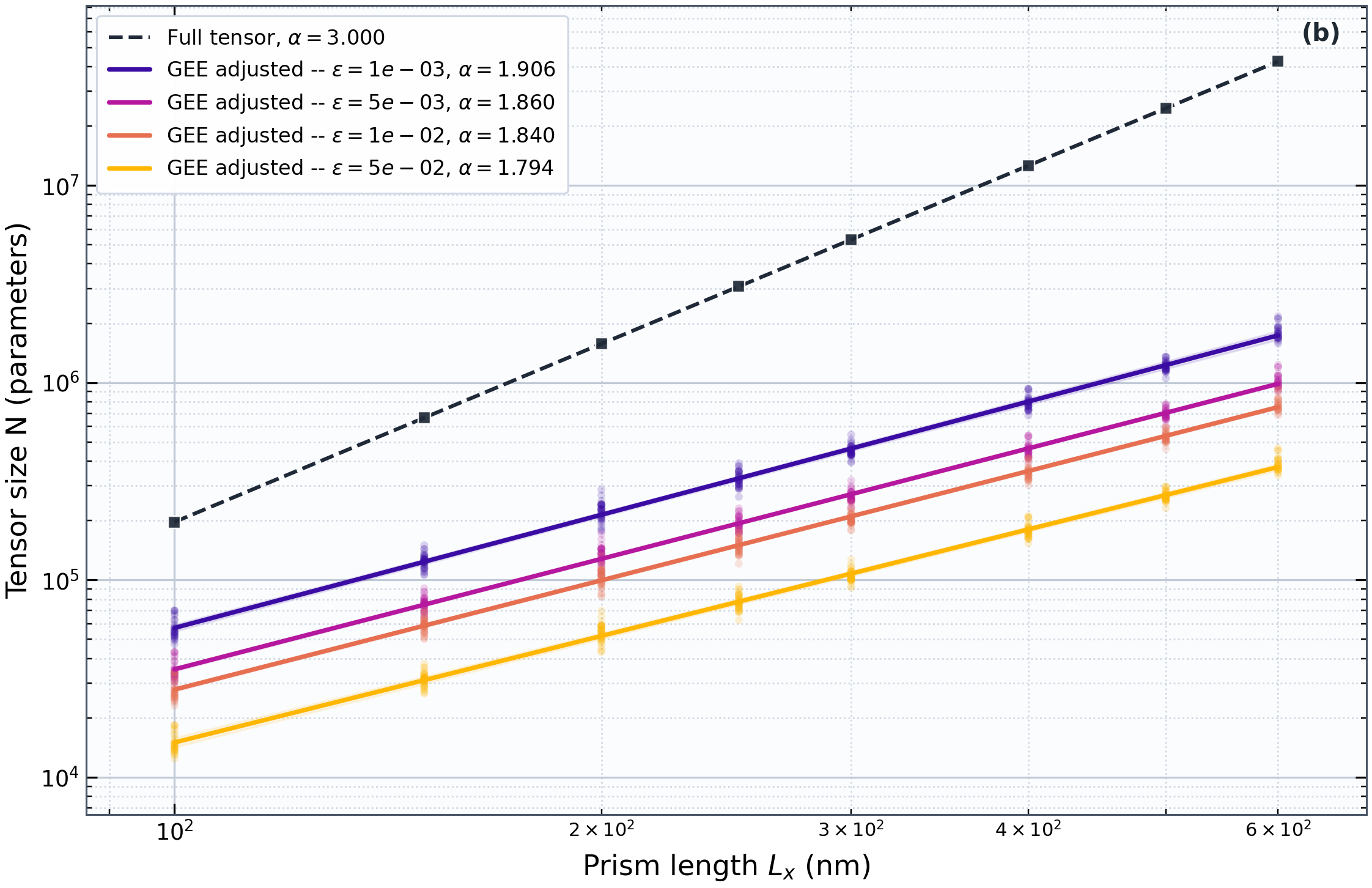}
  \caption{\textbf{Scaling as a function of the linear size (VS series).} (a) CR as a function of MPVE for the TT-TCI compression of flux-closure states in prisms of increasing length $L_x$, with fixed cell size $a=1.5625\,\mathrm{nm}$. We plot the mean variation and $90 \%$ CI envelope, estimated over five different initial pivots for the TCI.  Over the plotted error range, the CR monotonously {\em grows} with the prism size, showing that volumetric complexity does not control the TT scaling. (b) Regression-adjusted CPC versus $L_x$, obtained from all data in the VS series, over $10^{-3}\leq \mathrm{MPVE}\leq 5\times10^{-2}$, with power law fitted exponents $\alpha$. The original densely sampled field scales evidently as $L_x^3$, whereas the TT-compressed data scale as $L_x^{1.79\text{--}1.91}$. This is consistent with the TT-TCI compression taking full advantage of a complexity governed primarily by two-dimensional localized closure and wall structures rather than by the full 3D configuration.}
  \label{fig:size_panels}
\end{figure*}

\section{Results}
\label{sec:results}

The central question addressed by this work is whether the TT-compressed complexity of a three-dimensional micromagnetic equilibrium state is controlled by the full discretized volume, or instead by the lower-dimensional magnetic structures that carry most of the spatial variation. As introduced in Sec.~\ref{sec:methods_micromagnetic}, we considered a variable-size (VS) series to address this question directly, with the grid refinment being kept fixed while the prism is enlarged. A complementary variable-refinement (VR) series was also considered to isolates the issue of how the TT representation evolves under mesh refinement at fixed geometry.

\subsection{Scaling with object size}
\label{sec:results_vs}

For the VS series, the cell size is kept fixed at $a=1.5625\,\mathrm{nm}$, corresponding to $l_{\mathrm{ex}}/a\approx 3.638$, while the prism dimensions are increased at constant aspect ratio. Figure~\ref{fig:size_panels}(a) shows that the CR--MPVE curves move upward as the prism size increases. Over the plotted MPVE range, from $10^{-3}$ to $5\times10^{-2}$, larger prisms clearly allows for larger CR at any fixed MPVE. The regression-adjusted CPC scaling analysis reported in Fig.~\ref{fig:size_panels}(b) gives the fitted scaling exponents, for an assumed CPC versus linear dimension power law, as $1.906$ with a $95\%$ CI of $[1.864,1.948]$ at $10^{-3}$ MPVE, $1.860$ with $[1.844,1.876]$ CI at $5\times10^{-3}$ MPVE, $1.840$ with $[1.825,1.855]$ CI at $10^{-2}$ MPVE, and $1.794$ with $[1.753,1.835]$ CI at $5\times10^{-2}$ MPVE. Over the explored size range, the TT-TCI compressed data therefore grow slightly sub-quadratically with the linear dimension of the simulated magnetic prism, whereas the size of the densely discretized initial data grows evidently as $L^3$. This observed scaling is what is expected for a compression scheme taking full advantage of a spatially sparse information content dominated by structures of reduced dimensionality, namely the localized high-gradient walls and edge-turning regions that organize the flux-closure state. As the sample is enlarged, these structures expend as set of intersecting surface-like manifolds embedded in the 3D volume of the prism, while the volumetric domains remain nearly uniformly magnetized. The observed near-quadratic scaling is consistent with TCI derived TT representations tracking near optimally and selectively the growth of these high-gradient low-dimensional regions, rather than the volume of the full simulation domain. Consequently, the TT representation vastly outperforms the usual dense grid sampling, with a data structure size advantage increasing slightly faster than linearly with the simulated object size.

\begin{figure*}[!t]
  \centering
  \includegraphics[width=0.49\textwidth]{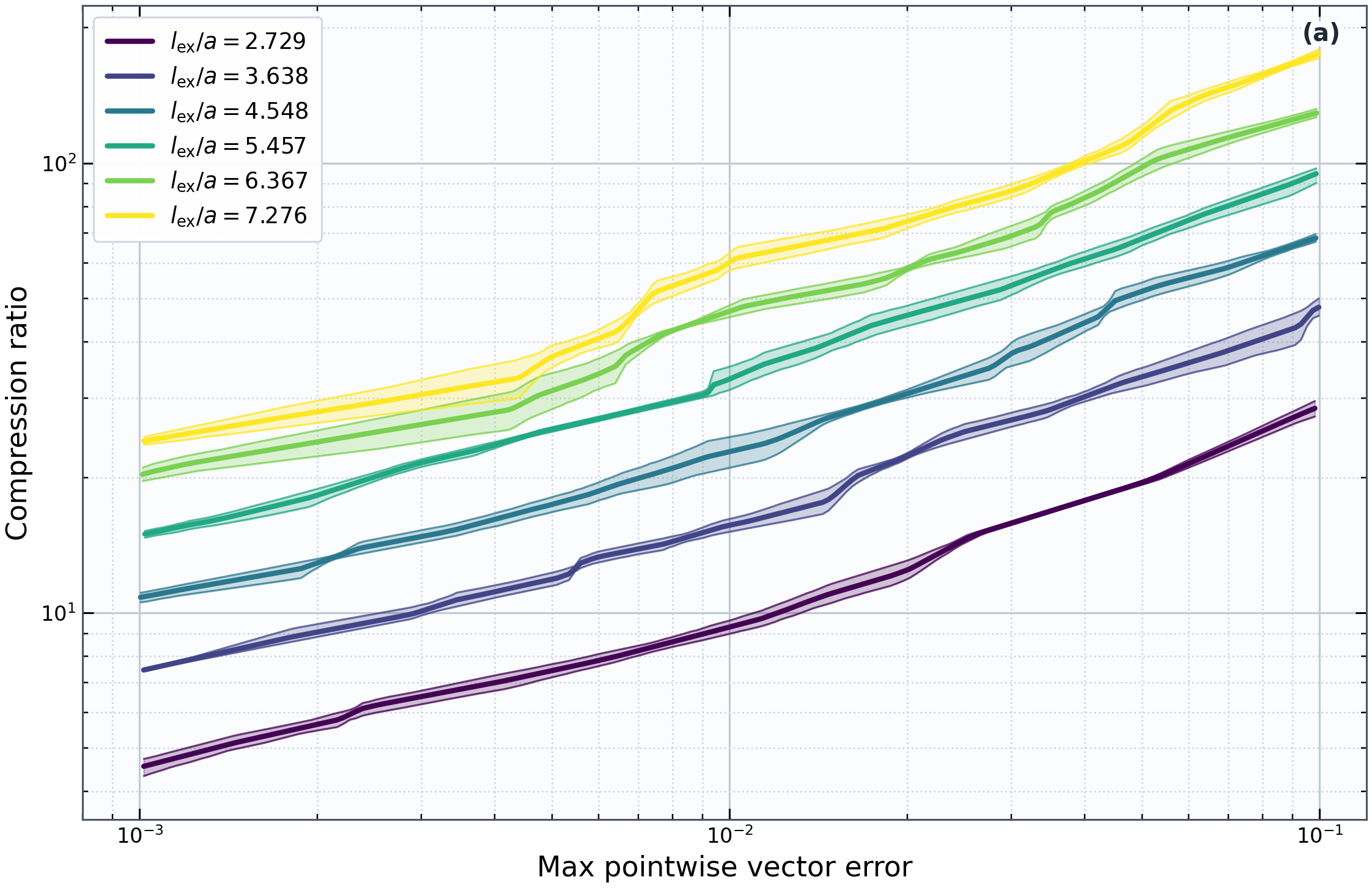}\hfill
  \includegraphics[width=0.49\textwidth]{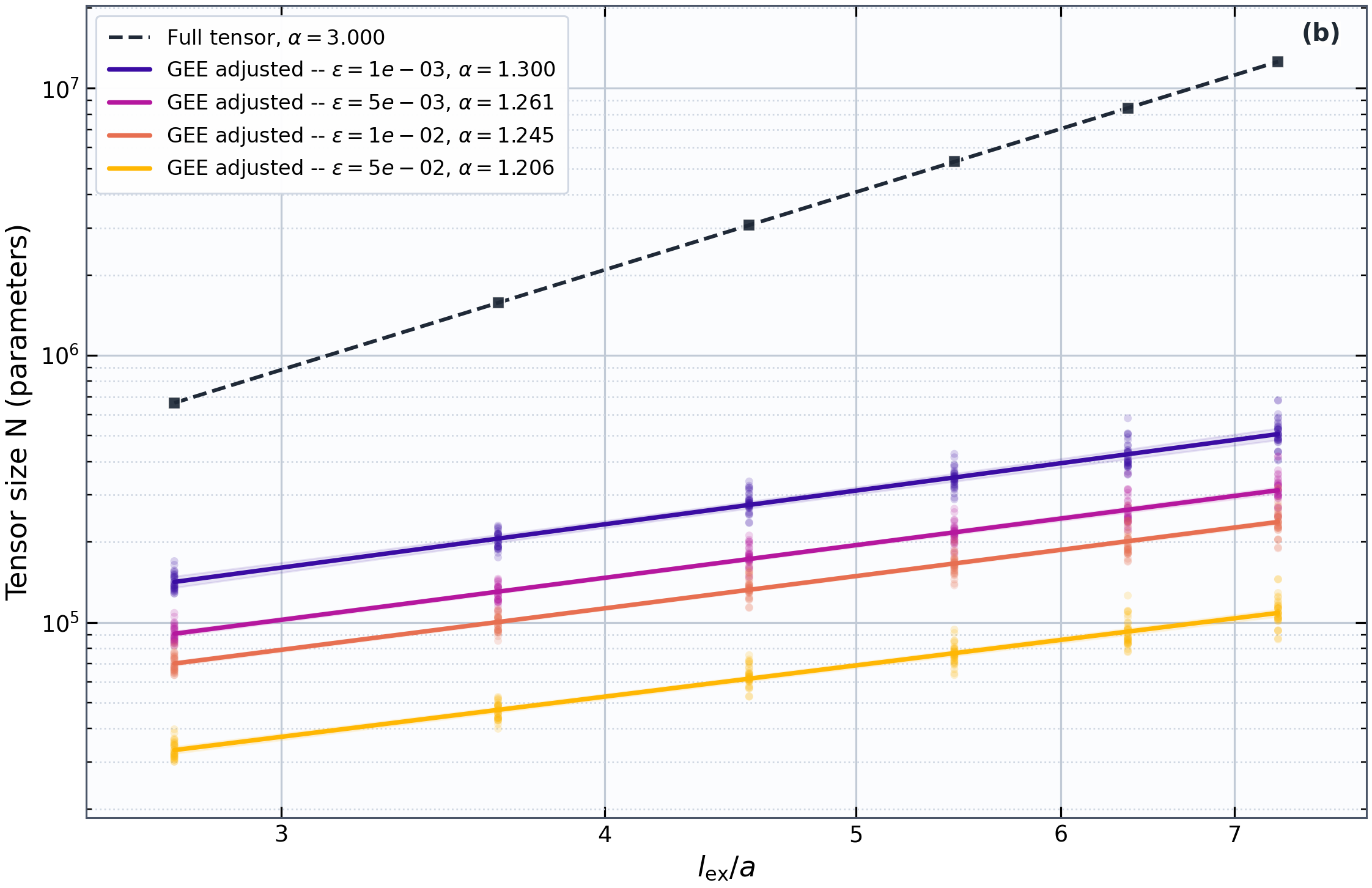}
  \caption{\textbf{Scaling as a function of grid refinement (VR series).} (a) CR as a function the MPVE for the TT-TCI compressed magnetization data of the $200\times100\times100\,\mathrm{nm}^3$ prism as the discretization is refined. The monotonous upward shift of the curves with increasing $l_{\rm ex} / a$ shows that finer resolution does not hurt compressibility; to the contrary, the TT representation captures the same physical configuration increasingly efficiently relative to the dense discretization. (b) Regression-adjusted CPC versus $l_{\mathrm{ex}}/a$, obtained from all in the VR series, over $10^{-3}\leq \mathrm{MPVE}\leq 5\times10^{-2}$. The fitted power law exponents remain in the range $1.21\text{--}1.30$, far slower than the $(l_{\mathrm{ex}}/a)^3$ scaling of the densely discretized field.}
  \label{fig:refinement_panels}
\end{figure*}

\subsection{Scaling with grid refinement}
\label{sec:results_fs}

In the VR series, the prism geometry is kept constant, and only the grid cell size $a$ is refined. The question which is addressed is therefore: once the same walls, closure regions, and edge-turning structures are already present at fixed physical locations, how does the compressed complexity grow when those existing structures are sampled more densely ? Figure~\ref{fig:refinement_panels}(a) shows that the CR--MPVE curves shift upward as $l_{\rm ex} / a$ increases, so at any given MPVE the refined fields become more compressible relative to a dense sampling, not less. In Fig.~\ref{fig:refinement_panels}(b), one sees that the regression-adjusted CPC trends remain far below cubic growth across the full plotted error window. The fitted power law scaling gives exponents equal to $1.300$ with a $95\%$ CI of $[1.218,1.381]$ at $10^{-3}$ MPVE, $1.261$ with $[1.220,1.302]$ CI at $5\times10^{-3}$ MPVE, $1.245$ with $[1.216,1.273]$ CI at $10^{-2}$ MPVE, and $1.206$ with $[1.162,1.250]$ CI at $5\times10^{-2}$ MPVE. This near-linear behavior has an obvious interpretation. At fixed physical size, refinement does not enlarge the lateral extent of the high-gradient regions; the walls, closure regions, and edge structures keep occupying the same locations and spanning the same physical area. What a sampling refinement adds is additional information primarily in the direction perpendicular to those mostly two-dimensional regions, that is, across their thickness. An optimal compressed representation should then acquire new degrees of freedom mainly along this wall-normal local direction. This is exactly the trend observed here. The fitted exponents remain slightly above unity probably because the active structures are not ideally two-dimensional, with the equilibrium state also containing edge-turning and vortex-related features. Nevertheless, the dominant scaling signal is clear: under grid refinement, the additional TT complexity is controlled mainly by increasing one-dimensional sampling normal to already-existing low-dimensional high-gradient structures, not by the growth in size of the full 3D mesh. Again, this vastly outperform the scaling of any near-uniform sampling which underpin traditional numerical micromagnetics methods, with a relative gain growing near quadratically with the grid refinement.

\section{Conclusion}
\label{sec:conclusion}

The newly established scaling laws for TT-TCI compression of simulated micromagnetic data are the central result of this work. Under physical-size growth at fixed spatial sampling, compressed TT representation grows near-quadratically. Under fixed-geometry refinement, it grows close to linearly. These two scaling laws probe different directions in problem size, yet they lead to the same conclusion: a compact TT representation can optimally take advantage of the inherent informational sparsity of micromagnetic configurations in the space domain, vastly outperforming traditional dense and nearly uniform sampling strategies. These newly established empirical results, with a clear conceptual interpretation, have implications beyond the specific study carried out here. It indicates that tensor-network methods are a computationally attractive route for large-scale micromagnetics, in which the solutions will be approximated in TT format and the operators in matrix product form. The natural next step is therefore the development of tensor-network based micromagnetic solvers. On the basis of our results, it appears that such future solvers have the potential to achieve the memory efficiency and computational performance usually associated with 1.5D methods while performing full fledged 3D simulations. This will open an entirely new era in computational micromagnetics, with a foreseeable far reaching impact in the fields of spintronics, magnonics and related applications.

\begin{acknowledgments}
One of the author (TV) acknowledges fruitful discussions with Xavier Waintal, Kei Yamamoto and Vincent Cros.
\end{acknowledgments}

\section*{Data Availability Statement}
The data that support the findings of this study are available from the corresponding author upon reasonable request.

\bibliography{references}

@article{de:2024,
  author  = {De, Saibal and Corona, Eduardo and Jayakumar, Paramsothy and Veerapaneni, Shravan},
  title   = {Tensor-train compression of discrete element method simulation data},
  journal = {Journal of Terramechanics},
  year    = {2024},
  volume  = {113-114},
  pages   = {100967},
  doi     = {10.1016/j.jterra.2024.100967},
  url     = {https://doi.org/10.1016/j.jterra.2024.100967}
}

@article{li:2025,
  author  = {Li, Keren and Zhang, Wenqiang and Xiao, Dandan and Hou, Peng and Yan, Shuai and Wang, Yang and Mao, Xuerui},
  title   = {Compression of big data collected in wind farm based on tensor train decomposition},
  journal = {Big Data Research},
  year    = {2025},
  volume  = {41},
  pages   = {100554},
  doi     = {10.1016/j.bdr.2025.100554},
  url     = {https://doi.org/10.1016/j.bdr.2025.100554}
}

@article{ritter:2024,
  title = {Quantics Tensor Cross Interpolation for High-Resolution Parsimonious Representations of Multivariate Functions},
  author = {Ritter, Marc K. and N\'u\~nez Fern\'andez, Yuriel and Wallerberger, Markus and von Delft, Jan and Shinaoka, Hiroshi and Waintal, Xavier},
  journal = {Phys. Rev. Lett.},
  volume = {132},
  issue = {5},
  pages = {056501},
  numpages = {6},
  year = {2024},
  month = {Jan},
  publisher = {American Physical Society},
  doi = {10.1103/PhysRevLett.132.056501},
  url = {https://link.aps.org/doi/10.1103/PhysRevLett.132.056501}
}

@article{oseledets:2011,
  title        = {Tensor-{Train} Decomposition},
  author       = {Oseledets, I. V.},
  journal      = {SIAM Journal on Scientific Computing},
  year         = {2011},
  volume       = {33},
  number       = {5},
  pages        = {2295--2317},
  doi          = {10.1137/090752286}
}

@article{oseledets:2010,
title = {TT-cross approximation for multidimensional arrays},
journal = {Linear Algebra and its Applications},
volume = {432},
number = {1},
pages = {70-88},
year = {2010},
issn = {0024-3795},
doi = {https://doi.org/10.1016/j.laa.2009.07.024},
url = {https://www.sciencedirect.com/science/article/pii/S0024379509003747},
author = {Ivan Oseledets and Eugene Tyrtyshnikov}
}

@article{dolgov:2020,
title = {Parallel cross interpolation for high-precision calculation of high-dimensional integrals},
journal = {Computer Physics Communications},
volume = {246},
pages = {106869},
year = {2020},
issn = {0010-4655},
doi = {https://doi.org/10.1016/j.cpc.2019.106869},
url = {https://www.sciencedirect.com/science/article/pii/S0010465519302565},
author = {Sergey Dolgov and Dmitry Savostyanov}
}

@article{mumax3,
  title        = {The design and verification of {MuMax3}},
  author       = {Vansteenkiste, Arne and Leliaert, Jonathan and Dvornik, Mykola and Helsen, Mathias and Garcia-Sanchez, Felipe and Van Waeyenberge, Bartel},
  journal      = {AIP Advances},
  year         = {2014},
  volume       = {4},
  number       = {10},
  pages        = {107133},
  doi          = {10.1063/1.4899186}
}

@article{Moreels2026,
  author       = {Moreels, Lars and Lateur, Ian and De Gusem, Diego and Mulkers, Jeroen and Maes, Jonathan and Milo\v{s}evi\'c, Milorad V. and Leliaert, Jonathan and Van Waeyenberge, Bartel},
  title        = {mumax+: extensible GPU-accelerated micromagnetics and beyond},
  journal      = {npj Computational Materials},
  year         = {2026},
  volume       = {12},
  number       = {1},
  pages        = {71},
  issn         = {2057-3960},
  doi          = {10.1038/s41524-025-01893-y},
  url          = {https://doi.org/10.1038/s41524-025-01893-y}
}

@article{UBS:22,
  author       = {Usvyatsov, Mikhail and Ballester-Ripoll, Rafael and Schindler, Konrad},
  title        = {tntorch: Tensor Network Learning with {PyTorch}},
  journal      = {Journal of Machine Learning Research},
  year         = {2022},
  volume       = {23},
  number       = {208},
  pages        = {1--6},
  url          = {http://jmlr.org/papers/v23/21-1197.html}
}

@article{paszke2019pytorch,
  title        = {PyTorch: An Imperative Style, High-Performance Deep Learning Library},
  author       = {Paszke, Adam and Gross, Sam and Massa, Francisco and Lerer, Adam and Bradbury, James and Chanan, Gregory and Killeen, Trevor and Lin, Zeming and Gimelshein, Natalia and Antiga, Luca and Desmaison, Alban and Kopf, Andreas and Yang, Edward and DeVito, Zachary and Raison, Martin and Tejani, Alykhan and Chilamkurthy, Sasank and Steiner, Benoit and Fang, Lu and Bai, Junjie and Chintala, Soumith},
  journal      = {arXiv preprint arXiv:1912.01703},
  year         = {2019},
  url          = {https://arxiv.org/abs/1912.01703}
}

@InProceedings{mckinney-proc-scipy-2010,
  author    = {Wes McKinney},
  title     = {Data Structures for Statistical Computing in Python},
  booktitle = {Proceedings of the 9th Python in Science Conference},
  pages     = {56--61},
  year      = {2010},
  editor    = {St\'efan van der Walt and Jarrod Millman},
  doi       = {10.25080/Majora-92bf1922-00a}
}

@inproceedings{seabold2010statsmodels,
  title     = {statsmodels: Econometric and statistical modeling with python},
  author    = {Seabold, Skipper and Perktold, Josef},
  booktitle = {9th Python in Science Conference},
  year      = {2010}
}

@techreport{oommf,
  title        = {{OOMMF} User's Guide, Version 1.0},
  author       = {Donahue, Michael J. and Porter, Donald G.},
  institution  = {National Institute of Standards and Technology},
  year         = {1999},
  number       = {NISTIR 6376}
}

@article{abert:2019,
  title        = {Micromagnetics and spintronics: models and numerical methods},
  author       = {Abert, Christian},
  journal      = {European Physical Journal B},
  year         = {2019},
  volume       = {92},
  pages        = {120},
  doi          = {10.1140/epjb/e2019-90568-2}
}

@misc{pisoni:2025,
  title        = {Compression, simulation, and synthesis of turbulent flows with tensor trains},
  author       = {Pisoni, Stefano and Peddinti, Raghavendra D. and Guzman, Siddhartha E. and Tiunov, Egor and Aolita, Leandro},
  year         = {2025},
  note         = {Manuscript; see source in the accompanying project directory}
}

@article{danis:2025,
title = {Tensor-train WENO scheme for compressible flows},
journal = {Journal of Computational Physics},
volume = {529},
pages = {113891},
year = {2025},
issn = {0021-9991},
doi = {https://doi.org/10.1016/j.jcp.2025.113891},
url = {https://www.sciencedirect.com/science/article/pii/S0021999125001743},
author = {M. Engin Danis and Duc Truong and Ismael Boureima and Oleg Korobkin and Kim {\O}. Rasmussen and Boian S. Alexandrov}
}

@article{fernandez:2017,
author  = {Fernandez-Pacheco, Amalio and Streubel, Robert and Fruchart, Olivier and Hertel, Riccardo and Fischer, Peter and Cowburn, Russell P.},
  title        = {Three-dimensional nanomagnetism},
  journal      = {Nature Communications},
  year         = {2017},
  volume       = {8},
  pages        = {15756},
  doi          = {10.1038/ncomms/15756}
}

@article{brown:1963,
  title        = {Micromagnetics},
  author       = {Brown, W. F.},
  journal      = {Interscience Publishers},
  year         = {1963},
  volume       = {},
  number       = {},
  pages        = {},
  doi          = {}
}

@article{trabzon:2025,
author  = {Trabson, Ahmet Bahadir and Goto, Taichi and Onsbali, Mehmet Cengiz},
  title        = {Comprehensive Micromagnetic Modeling: Practical Techniques, Applications and Emerging Challenges},
  journal      = {IEEE Access},
  year         = {2025},
  volume       = {13},
  pages        = {93102},
  doi          = {10.1109/ACCESS.2025.3573946}
}

@article{yan:2007,
author  = {Yan, Ming and Hertel, Riccardo and Schneider, Claus M.},
  title        = {Calculations of three-dimensionnal magnetic normal modes in mesoscopic permalloy prism with vortex structure},
  journal      = {Phys. Rev. B},
  year         = {2007},
  volume       = {76},
  pages        = {094407},
  doi          = {10.1103/PhysRevB.76.094407}
}

@article{gu:2025,
author  = {Gubbiotti, Gianluca and al.},
  title        = {2025 Roadmap in 3D nanomagnetism},
  journal      = {J. Phys. Condens.Matter.},
  year         = {2025},
  volume       = {37},
  pages        = {143502},
  doi          = {10.1088/1361-648X/ad9655.PHID:39577093}
}

\end{document}